# Double-pulse control of all optical magnetization reversal in Tb/Co multilayers

Sheng Li[1,2*], Dinar Khusyainov[2], Rein Liefferink[2], Lukas Körber[2], Quoc-Trung Trinh[3], Ricardo C. Sousa[3], Liliana D. Buda-Prejbeanu[3], Johan H. Mentink[2], Theo Rasing[2], and Alexey V. Kimel[2]

*1 International School of Materials Science and Engineering, Wuhan University of Technology, Wuhan 430070, China*
*2 Radboud University, Institute for Molecules and Materials, Nijmegen 6525 AJ, the Netherlands*
*3 Université Grenoble Alpes, CEA, CNRS, Grenoble INP, SPINTEC, 38000 Grenoble, France*

*Corresponding author. E-mail: s.li@whut.edu.cn

**Abstract:** Recent experiments have shown that femtosecond laser pulse with a Gaussian intensity profile can induce magnetization reversal in Tb/Co multilayers with a ring-shaped switching pattern within the laser-irradiated area. Here, we investigate the ultrafast magnetization dynamics leading to such a ring-shaped switching by using double-pulse laser excitation. The laser pulses cause heat-induced quenching and subsequent recovery of the magnetic anisotropy in the multilayers and drive the precessional magnetization switching in the magnetic multilayers. By adjusting the delay between the two pump pulses, we demonstrate that the recovery process can be manipulated and show, experimentally and numerically, that this allows control over the final magnetic domain pattern.

## 1. Introduction

With the rapid progress of artificial intelligence, there is a growing demand for faster and energy-efficient information processing and storage. Ultrafast laser pulses are the shortest physical stimuli known to trigger magnetization dynamics [1], which led to the discovery of all-optical magnetization reversal [2-5] and gave rise to the field of ultrafast magnetism [6-8]. Helicity-dependent all-optical switching (HD-AOS), which relies on the polarization of laser pulses, has been demonstrated in both metallic and dielectric thin films [2, 9-11]. This phenomenon can be attributed to the differences in absorption of circularly polarized light with opposite helicity, depending on the orientation of the magnetization [12]. Furthermore, AOS can also be induced without any specific requirements regarding the polarization of the light [3, 13-15]. This type of polarization-independent AOS is typically observed in ferrimagnetic systems, where laser-induced switching is driven by an exchange-mediated angular momentum transfer between two collinear but oppositely magnetized sublattices [4, 16].

In addition to the conventional AOS scenarios mentioned above, laser-induced magnetization reversal under an in-plane magnetic field can lead to precessional switching, often accompanied by ring-shaped domain patterns [17-19]. Davies et al. [17] demonstrated that precessional reversal of magnetization in yttrium iron garnet can be achieved by thermal modulation of the magnetocrystalline anisotropy, where the interplay between magnetic precession and anisotropy recovery results in inhomogeneous, ring-shaped, switching patterns by Gaussian laser pulses. Similar phenomena have also been observed in a Pt/Co/Pt trilayer under an in-plane magnetic field [18]. However, other experiments conducted on synthetic ferrimagnetic multilayers have revealed single-pulse spatially switched rings occurring without the requirement of external magnetic fields [20-24]. The thermal modulation of the anisotropy by the laser pulse [22], combined with a tilt of the anisotropy axis [21], provides a possible explanation for this type of field-free switching. Previous studies have demonstrated that the switching behavior can be tuned by modifying the multilayer composition [24]. It is also of interest to explore whether alternative routes could be used to control the switching patterns.

To clarify the mechanism of field-free ring-shaped switching and explore a potential new route of dynamical control, we investigated laser-induced magnetization reversal in a ferrimagnetic $[Tb/Co]_5$ multilayer using a double-pulse approach, which involves changing the time interval between the two laser pulses. By tuning the time interval and varying their fluences, we observed controllable transformations between switching patterns with and without rings. Using numerical simulations based on the Landau-Lifshitz-Gilbert equation (LLG) combined with a microscopic three-temperature model (M3TM), we obtained the expected patterns considering a heat-induced modulation of the magnetic anisotropy. The experimental results are consistent with the numerical simulations, validating the hypothesis and confirming that the double-pulse approach is a simple way to control the magnetization configuration in $[Tb/Co]_5$ multilayers.

## 2. Experimental methods

The sample analyzed in this study is a synthetic ferrimagnetic $[Tb/Co]_5$ multilayer deposited by dc magnetron sputtering on a silicon wafer substrate. The Tb and Co layers are grown in wedges with independent spatial dependences of their thicknesses. The experiments were performed at a position with the composition $[Tb(0.6\ nm)/Co(1.4\ nm)]_5$, which is located in the Co-dominant composition exhibiting toggle-switching behavior (see **Supplemental Note 1**), consistent with previous reports [20, 21, 23]. The experiments are performed in double-pulse excitation and static probe regimes. The pump laser pulses are generated from an amplified Ti:Sapphire laser system with a central wavelength of 800 nm and a pulse duration of roughly 500 fs at the sample position. As sketched in **Fig. 1a**, two consecutive linearly polarized laser pulses with a controlled time delay $\Delta\tau_{1-2}$ between each other using an optical delay stage, are focused onto the multilayer stack at an incident angle ~ 30°. The two pulses are set with orthogonal polarization to exclude the influence of spatial-temporal interference. The focused spot radius at FWHM for Pump1 and Pump2 are slightly different and calibrated via the Liu method [25] to be $44\times32\ \mu m^2$ and $38\times32\ \mu m^2$, respectively. The static magnetic domain structures are observed by using magneto-optical Kerr (MOKE) microscopy. An electromagnet is used to saturate and reset the magnetization of the sample. For image analysis, the images obtained from the laser excitation measurement are normalized with the background images and processed using a Gaussian filter to improve visualization.

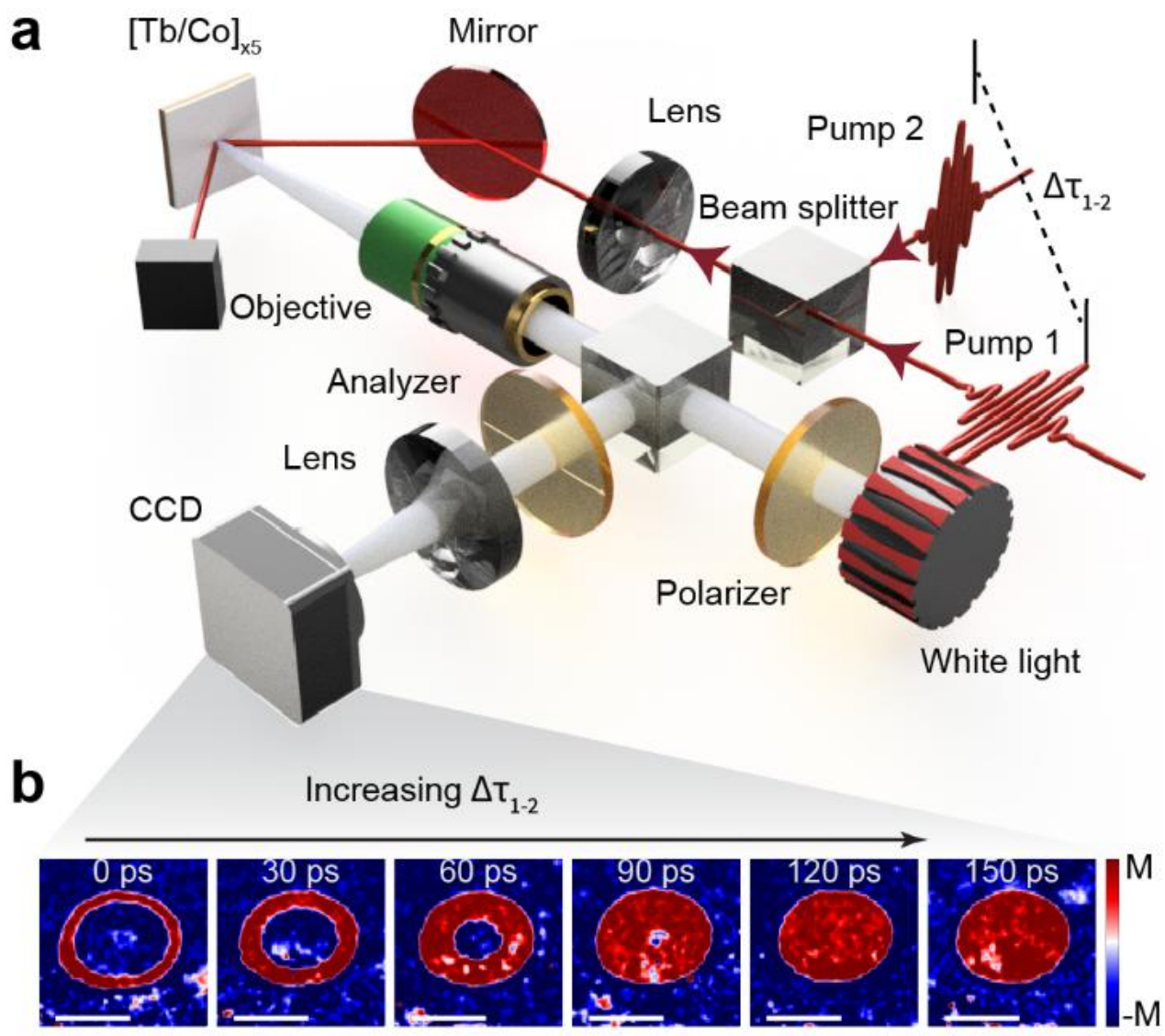

**FIG. 1. (a)** Schematic of the experimental system for double-pulse excitation and static magneto-optical Kerr imaging. **(b)** Magnetic domain structures recorded after double-pulse excitation showing the gradual evolution of switched rings into fully switched spots upon increasing the time delay $\Delta\tau_{1-2}$. The fluence values of the first and the second pump are 4.5 mJ/cm$^2$ and 5.4 mJ/cm$^2$, respectively. The scale bars are 50 μm.

## 3. Results and discussions

First, by varying $\Delta\tau_{1-2}$ between 0 ps and 150 ps, we observed the formation of the magnetic rings and their gradual transformation into fully switched spots as the pulses are further separated in time (see **Fig. 1b**). This suggests that this type of AOS differs from ultrafast toggle switching, which relies on longitudinal angular momentum transfer in typical ferrimagnetic systems. There, the magnetization dynamics is usually completed within a few tens of picoseconds, so that a second pulse reverses the magnetization back within one hundred picoseconds [26, 27]. For comparison, Wang et al. [28] reported that, in ferrimagnetic $Gd_{27}Fe_{63.87}Co_{9.13}$ alloy, the time separation between two pump pulses has little influence on the final magnetic state, and the second pulse only begins to induce back switching when the separation is increased to several hundreds of picoseconds. However, here we find that even slight adjustments in the delay significantly change the final magnetic pattern.

According to a previous study [22], in this type of thermally driven AOS, the local heating affects the extent of anisotropy quenching, which will define the initial torque exerted on the magnetization. Therefore, we studied the fluence dependence for dynamical double-pulse control of AOS with smaller steps of $\Delta\tau_{1-2}$ in [Tb(0.6 nm)/Co(1.4 nm)]$_5$ (**Fig. 2**) to obtain more information about the dynamics of the switching. In **Fig. 2a** and **Fig. 2b**, two columns display the magnetization reversal separately induced by Pump1 and Pump2 at different fluences. For these single-pulse excitation, the AOS occurs within a certain fluence window with ring structures appearing at high fluence, demonstrating a threshold effect. The initial torque that triggers this spin reorientation can be provided either by an externally applied in-plane magnetic field [17-19] or by the intrinsic canting of the magnetocrystalline anisotropy away from the out-of-plane direction [21]. In **Fig. 2c**, combining these two pulses, a diagram collecting the MOKE images versus both laser fluence and $\Delta\tau_{1-2}$ is presented. For a given pair of fluences, increasing the time delay between the two pump pulses changes the switched rings patterns into a homogeneously reversed magnetic domain. This behavior closely resembles the trend observed when decreasing the fluence at a constant time delay. At high fluence and small $\Delta\tau_{1-2}$, a second narrow AOS window also appears as observed using a single pulse (**Supplemental Note 1**). It suggests that increasing $\Delta\tau_{1-2}$ for double-pulse pumping has similar effects as decreasing the fluence for single-pump switching. Note that by employing double-pulse pumping, we can generate homogeneous spots with larger diameters that are not achievable using a single laser pulse.

Previous studies implied that the laser-induced switching process can be understood as a consequence of the interplay between spin reorientation and the subsequent recovery of the quenched local anisotropy [17, 22]. Under single-pulse excitation, it should only depend on the initial rate of anisotropy quenching, while the recovery process remains uncontrolled. Using two excitation pulses, we demonstrate that we can control this recovery process, i.e. the temporal evolution of the laser-induced anisotropy changes and show, experimentally and numerically, that this can lead to a control of the switched domain structure.

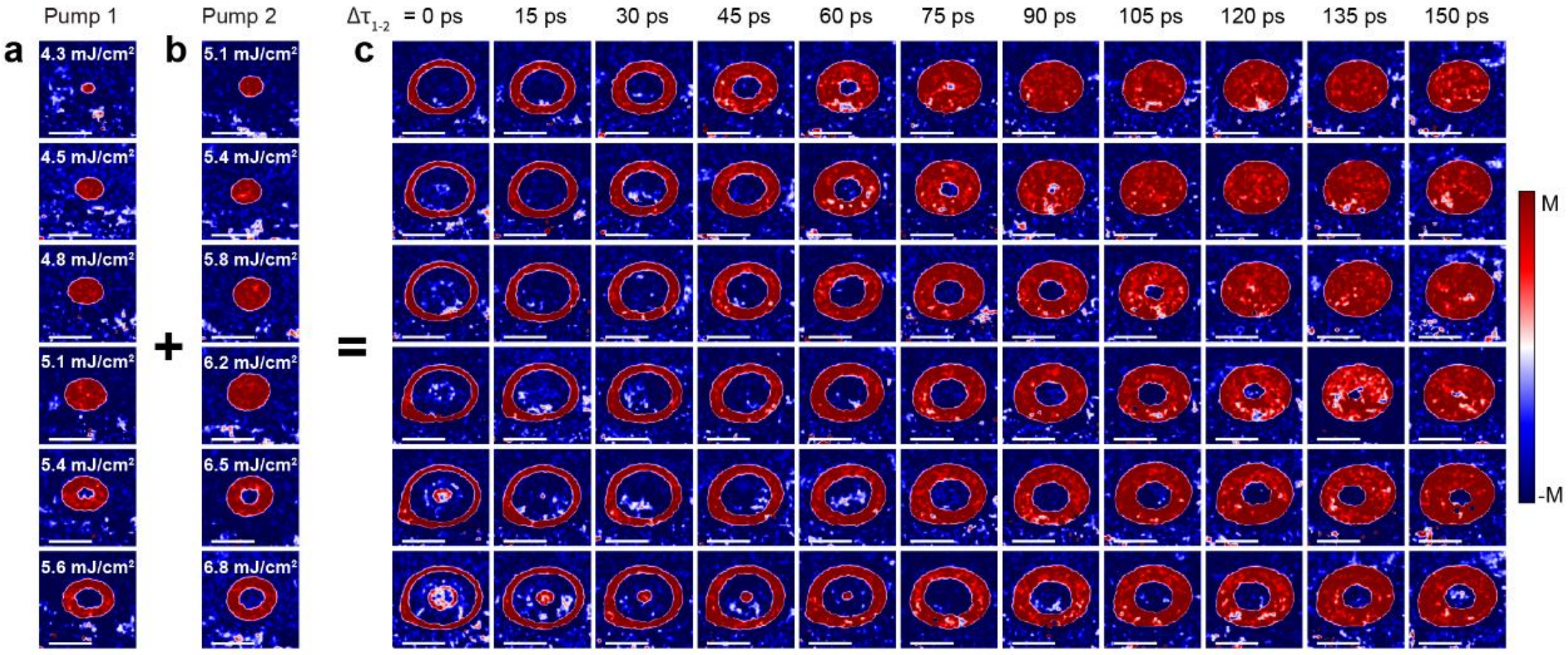

**FIG. 2.** Single-pulse MOKE images obtained using only pump 1 (**a**) and only pump 2 (**b**) while varying the fluence. **(c)** Double-pulse MOKE images obtained by combining Pump1 and Pump2 when varying both fluence and time delay $\Delta\tau_{1-2}$. The pixels in each image are normalized with two background images. The scale bars are 50 μm.

To describe the above-mentioned scenario, we performed simulations based on the Landau Lifshitz-Gilbert equation (LLG):

$$\frac{d\mathbf{m}}{dt} = -\frac{\gamma}{1+\alpha^2}[\mathbf{m}\times\mathbf{B}_{\text{eff}} + \alpha\mathbf{m}\times(\mathbf{m}\times\mathbf{B}_{\text{eff}})], \quad (1)$$

describing damped precessional magnetization dynamics where $\gamma$, $\alpha$, $\mathbf{m}$, $\mathbf{B}_{\text{eff}}$ are the gyromagnetic ratio, magnetic damping, unitary magnetization vector, and effective magnetic field, respectively. The effective field $\mathbf{B}_{\text{eff}}$ equals $-\frac{1}{M}\frac{\partial E}{\partial \mathbf{m}}$ in which the free energy density $E$ is given by:

$$E = -K_u(\mathbf{u}_k \cdot \mathbf{m})^2 + \frac{1}{2}\mu_0 M^2 m_z^2, \quad (2)$$

where $M$ is the magnetization amplitude, $K_u$ is the magnetocrystalline anisotropy and $\mathbf{u}_k = (\sin(\theta_u),0,\cos(\theta_u))$ is the axis of the anisotropy. A slight tilt of the anisotropy axis by an angle $\theta_u$ from the normal is introduced to ensure non-zero initial torques as used in ref. [21]. The assumption is supported by hysteresis measurements [21, 29]. The angle $\theta_u$ adopted in this simulation was set to 3°, where the simulated fluence dependence of the switching shows comparable trends to the experimental results.

Although the constant $K_u$ is commonly described as a function of temperature [21, 30], it can also be expressed as a function of $M$ according to the Callen-Callen power law [31]. The temperature dependence arises from the magnetization change induced by thermal fluctuations and associated entropy effects. Therefore, we adopted the Callen-Callen relation for materials with uniaxial anisotropy as follows [31]:

$$\frac{K_u(M(t))}{K_u(0)} = \left[\frac{M(t)}{M(0)}\right]^3, \quad (3)$$

Based on this assumption, we have investigated the effect of the laser-induced thermal modulation of the anisotropy on the magnetization dynamics based on the evolution of magnetization amplitude $M$(t). To achieve this, we use the microscopic three-temperature model (M3TM) [32], which describes the ultrafast demagnetization by modeling the coupled energy and angular momentum transfer among electrons, lattices and spins:

$$C_e \frac{dT_e(t)}{dt} = -G_{el}[T_e(t) - T_l(t)] + P(t), \quad (4)$$

$$C_l \frac{dT_l(t)}{dt} = -G_{el}[T_l(t) - T_e(t)] - \frac{C_l}{\tau_r}[T_l(t) - T_0], \quad (5)$$

$$\frac{d\eta(t)}{dt} = R\eta(t)\frac{T_l}{T_c}\left[1 - \eta(t)\coth\left(\frac{\eta(t)T_c}{T_e}\right)\right], \quad (6)$$

where $T_e$, $C_e$, $T_l$, $C_l$ correspond to the effective temperature and heat capacity of the electron and lattice systems, respectively; $T_c$ is the Curie temperature; $G_{el}$ is the coupling constant between the two systems. The relaxation term with the cooling rate $\tau_r$ in Equation (5) helps to drive the system back to the initial temperature, which is also taken into account in previous work for studying laser-induced magnetization dynamics caused by thermal modulation of the anisotropy in similar ferrimagnetic multilayers [21, 33, 34]. $P(t)$ is the temporal energy input by the ultrafast laser pulses. The evolution of $M$(t) is determined by Equation (6) where $\eta(t)$ equals $M$(t)/$M$(0). The factor $R$ is a material-specific demagnetization rate. Here, the thermal process is considered to be purely local, neglecting both exchange interactions between neighboring magnetic moments and spatio-temporal heat diffusion effects. Note that, by comparing the SQUID-VSM measurement for samples with different compositions in ref. [23], it is known that the compensation temperature of our ferrimagnetic multilayers should be far below room temperature in cobalt-dominant regions. Therefore, the system can be reasonably approximated as an effective ferromagnet within a macrospin model using a mean-field Weiss approach. Besides, the two pump pulses are modeled with identical spot sizes and fluences for computational convenience. The modeling parameters and details are provided in **Supplemental Note 2**.

The simulation results of the complete phase diagram of static double-pulse switching patterns are presented in **Supplemental Note 3**. It shows a very similar trend to the experimental results. This good agreement between simulation and experiments suggests that the hypothesis of local quenching of magnetic anisotropy is very likely to be valid. To better visualize this, we present vertically the cross-sections of the images showing the evolution of both experimental and modeling AOS results as a function of $\Delta\tau_{1-2}$ in **Fig. 3**. Under comparable fluence, both simulations and experiments show the trend from rings to homogeneous switched patterns. With increased $\Delta\tau_{1-2}$, the initially unswitched inner region disappears, while the outer switched ring gradually expands towards the center into homogeneous spots.

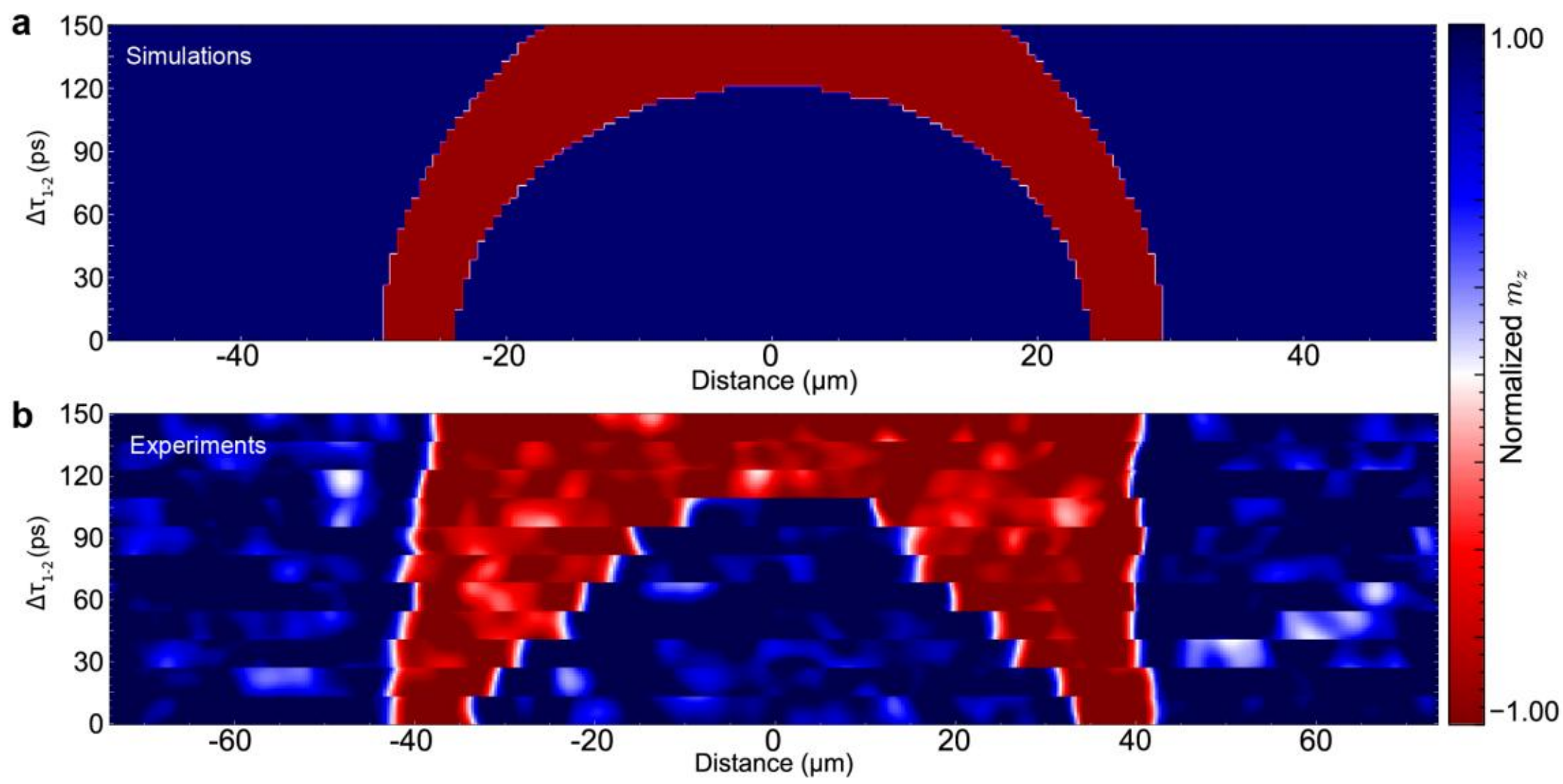

**FIG. 3.** The reversed domain patterns as a function of $\Delta\tau_{1-2}$ **(a)** simulations and **(b)** Experiments. The experimental plots are vertical stacks of cross-section rows in magneto-optical images with fluences of Pump1 4.8 mJ/cm$^2$, and Pump2 5.8 mJ/cm$^2$. For the simulations, the two pump pulses have the same shape and the same fluence of 4.4 mJ/cm$^2$. The color bar represents the normalized magnetization.

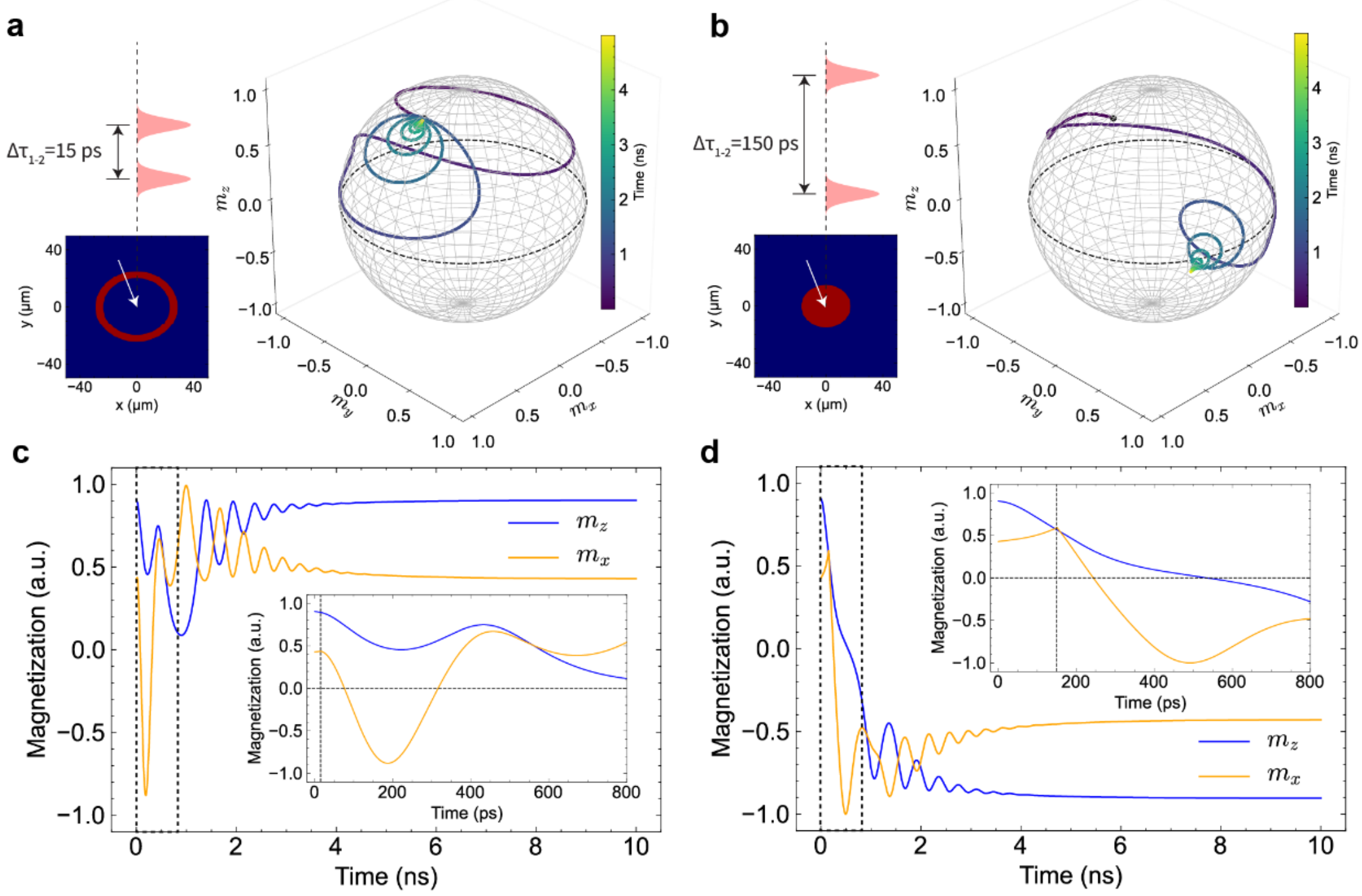

**FIG. 4.** The simulated magnetization patterns and three-dimensional magnetization trajectories after a double-pulse excitation with the same fluence 4.4 mJ/cm$^2$ for two pump pulses and different $\Delta\tau_{1-2}$ **(a)** 15 ps, **(b)** 150 ps, and the corresponding $m_x$, $m_z$ time traces for $\Delta\tau_{1-2}$ **(c)** 15 ps, **(d)** 150 ps. The dashed circles in (a) and (b) represent the plane of $m_z$=0. The color bars indicate the time delay after the first pump pulse. Zoomed-in views of the dynamics within the dashed rectangles in (c) and (d) are shown as insets, where the vertical dashed line indicates the timing of the second pulse.

From the agreement between the experimental and computational switching patterns, the magnetization dynamics of the strongly inhomogeneous, ring-like laser-induced switching in Tb/Co multilayers that are induced by a double-pulse pump can be studied via the simulation model. **Fig. 4a** and **Fig. 4b** show the trajectories of the magnetization simulated at the position of the beam center (see insets in **Fig. 4a** and **Fig. 4b**). We also plot the corresponding time traces in **Fig. 4c** and **4d** (see **Supplemental Note 4** for dynamics of $M$(t) and $K_u$(t)). In the simulation, there are long-lived oscillations of magnetization, while they are absent in the experiments [22, 23], which could come from averaging the signals over the whole spatial area of the probe, and also be possibly related to the transient increase in magnetic damping during the in-plane reorientation process [17]. Under a double-pulse laser fluence of 4.4 mJ/cm$^2$, no switching is obtained for a small interpulse delay ($\Delta\tau_{1-2}$ = 15 ps). The reversal of magnetization occurs when the two pulses are temporarily separated by 150 ps. In both cases, the magnetization undergoes large-amplitude in-plane precessions at the early stage when the anisotropy is suddenly quenched by the first pump. After this ultrafast stimulus, the magnetization is sensitive to changes of $\mathbf{B}_{\mathrm{eff}}$ and attracted to one of the metastable equilibrium directions once the anisotropy is partially recovered. Under moderate fluences, if the second pulse arrives on the sample at a short delay with respect to the first one, it leads to a strong degree of quenching and therefore accelerates the in-plane reorientation process by providing a larger initial torque. This could push the magnetization to cross the energy barrier that is only slightly recovered, and return back to the initial state. On the other hand, if the interpulse delay becomes larger, it will lower the degree of anisotropy quenching and allow more time for the energy barrier to rebuild after the second pulse. Therefore, it creates a threshold that prevents the magnetization to switch back, resulting in a reversed magnetization. This leads to a reversed magnetic pattern extended at large spatial configuration. On the other hand, at higher fluences that enable more periods of oscillation, the control from switched states to unswitched states can similarly be achieved. For instance, in the last row of the diagram in **Fig. 2** it is shown that the second switched spot in the center gradually shrinks and finally vanishes as the pulse delay is increased.

Our results show a dynamical double-pulse control of in-plane reorientational type of AOS in a ferrimagnetic multilayer. The switching was demonstrated to be a thermal process triggered by temporal shaping of the magnetic anisotropy using ultrafast heating [22]. Our macrospin model provides good insight into how switching occurs under double-pulse excitation by temporarily tailoring the magnetic anisotropy. By tuning the time separation between two pump pulses, our experimental results demonstrate a monotonous change of the switching patterns, providing a "coherent" control of the switching pattern. To compare, Zalewski et al. [35] recently reported the double-pulse coherent control of AOS in yttrium iron garnet, based on a non-thermal photomagnetic effect [5]. It was shown that the constructive interference of magnetic precession at a certain double-pulse delay can enhance the switching. The key idea of coherent control is the stabilization of the initial phase for ultrafast excitation, which is not expected for a thermal process. However, although the switching here is induced by a thermal quenching of magnetic anisotropy, the ultrafast time scale of this process compared to that of the precessional dynamics, leads to this apparent coherent control.

In a previous study, the consideration of constant magnetization enhances the shape anisotropy for driving the in-plane reorientation process [21]. Here, we show that even with quenched magnetization, it is still possible to achieve field-free precessional switching, provided that the reduction of the uniaxial anisotropy ($\propto M^3$ by Callen-Callen law) is more than that of the shape anisotropy ($\propto M^2$). It also implies that the field-free switching can be possibly tuned by tailoring the dependence of $K\mathrm{u}(M)$. One more assumption that deserves more attention is the hypothesis of a tilted axis of anisotropy. This hypothesis implies that the switching behavior should be sensitive to an applied in-plane bias field. To verify this, we performed field-dependent single-pulse switching experiments. In **Supplemental Note 5**, it is shown that the AOS is enhanced by applying an in-plane bias field in an arbitrary direction and suppressed in an opposite field. If no initial tilt would exist, the field-dependent switching would not rely on the direction of the applied field. Further experiments are required to clarify the origin of this tilt. In addition, further time-resolved measurements of the in-plane components of the magnetization in longitudinal or transverse Kerr configurations would be beneficial to understand the in-plane reorientation process and check the effective damping from experimental points of view.

## 4. Conclusions

In conclusion, we studied double-pulse AOS in a ferrimagnetic $[Tb/Co]_5$ multilayer. By using a double-pulse approach and numerical simulations, we demonstrate the control of the ring-shaped switching to homogeneous switching patterns using the time separation between the excitation pulses as a controlling parameter. These results provide more evidence that could help resolve the puzzle of field-free magnetization reversal with ring shapes. The effect of double-pulse excitation is revealed to be a time-dependent thermally induced change of the local magnetic anisotropy. This combined strategy appears to be a powerful tool for exploring ultrafast magnetization dynamics, while the method of multi-pump excitation to shape the time evolution of magnetic parameters offers additional degrees of freedom in controlling laser-induced switching patterns.

## Acknowledgements

The authors are grateful to K. Saeedi and C. Berkhout for their technical support. This work was funded by the Netherlands Organization for Scientific Research (NWO) and the European Research Council ERC Grant Agreement no. 856538 (3D-MAGiC). S. L. acknowledges the support from China Scholarship Council (CSC) fellowship no. 202306950107. L. K. acknowledges the funding from the Radboud Excellence Initiative and Innovation Programme Horizon Europe under grant agreement no. 101070290 (NIMFEIA). J.H.M acknowledges funding from the VIDI project no. 223.157 (CHASEMAG) and KIC project no. 22016 which are (partly) financed by the Dutch Research Council (NWO), as well as support from the European Union Horizon 2020 and innovation program under the European Research Council ERC Grant Agreement No. 856538 (3D-MAGiC) and the Horizon Europe project no. 101070290 (NIMFEIA). The samples in the paper were fabricated at PTA (Upstream Technological Platform) cleanroom which is funded from the French RENATECH network. The authors also acknowledge the France 2030 government grants ANR PEPR EMCOM (ANR-22-PEEL-009) for the financial support of this work.

## Author contributions

T. R. and A. V. K. conceived the project. D. K., T. R. and A. V. K. supervised the study. S. L. and D. K. built the setup for double-pulse experiments, and performed the experiments at Radboud University. S. L. carried out the numerical simulations with help from R. L., L. K., and J. H. M.. Q. T. T., R. C. S., L. D. B. P fabricated the sample and provided helpful guidance during the initial exploration of this project.

**Supplemental Materials for "Double-pulse control of all optical magnetization reversal in Tb/Co multilayers"**

Sheng Li[1,2*], Dinar Khusyainov[2], Rein Liefferink[2], Lukas Körber[2], Quoc-Trung Trinh[3], Ricardo C. Sousa[3], Liliana D. Buda-Prejbeanu[3], Johan H. Mentink[2], Theo Rasing[2], and Alexey V. Kimel[2]

*1 International School of Materials Science and Engineering, Wuhan University of Technology, Wuhan 430070, China*
*2 Radboud University, Institute for Molecules and Materials, Nijmegen 6525 AJ, the Netherlands*
*3 Université Grenoble Alpes, CEA, CNRS, Grenoble INP, SPINTEC, 38000 Grenoble, France*

*Corresponding author. E-mail: s.li@whut.edu.cn

## Contents:

## 1. Experimental details and sample information

The sample used in this study is a $[Tb/Co]_5$ synthetic ferrimagnetic multilayer deposited by dc magnetron sputtering with an annealing temperature 250 °C on a silicon wafer, showing perpendicular magnetic anisotropy and toggle switching phenomena. The substrate under $[Tb/Co]_5$ multilayer consists of a magnetic tunnel junction structure and a silicon wafer (**Fig. S1a**). The terbium and cobalt layers are sputtered as wedges on a wafer, which results in a spatially varying magnetic phase diagram depending on the compositions similar to ref. [1,2]. Here, all experiments were performed in roughly the same position with composition $[Tb(0.6\ nm)/Co(1.4\ nm)]_5$ where cobalt is dominant. In **Fig. S1b**, the hysteresis for $[Tb(0.6\ nm)/Co(1.4\ nm)]_5$ is obtained by averaging pixel values in MOKE images under different out-of-plane fields. The square hysteresis loop indicates the out-of-plane magnetic anisotropy. With the composition $[Tb(0.6\ nm)/Co(1.4\ nm)]_5$, the magnetization can be deterministically switched by single femtosecond laser pulse and toggled back by the next pulse following laser repetition (**Fig. S1c**). Between 5 mJ/cm$^2$ and 10 mJ/cm$^2$, two switching ranges are observed.

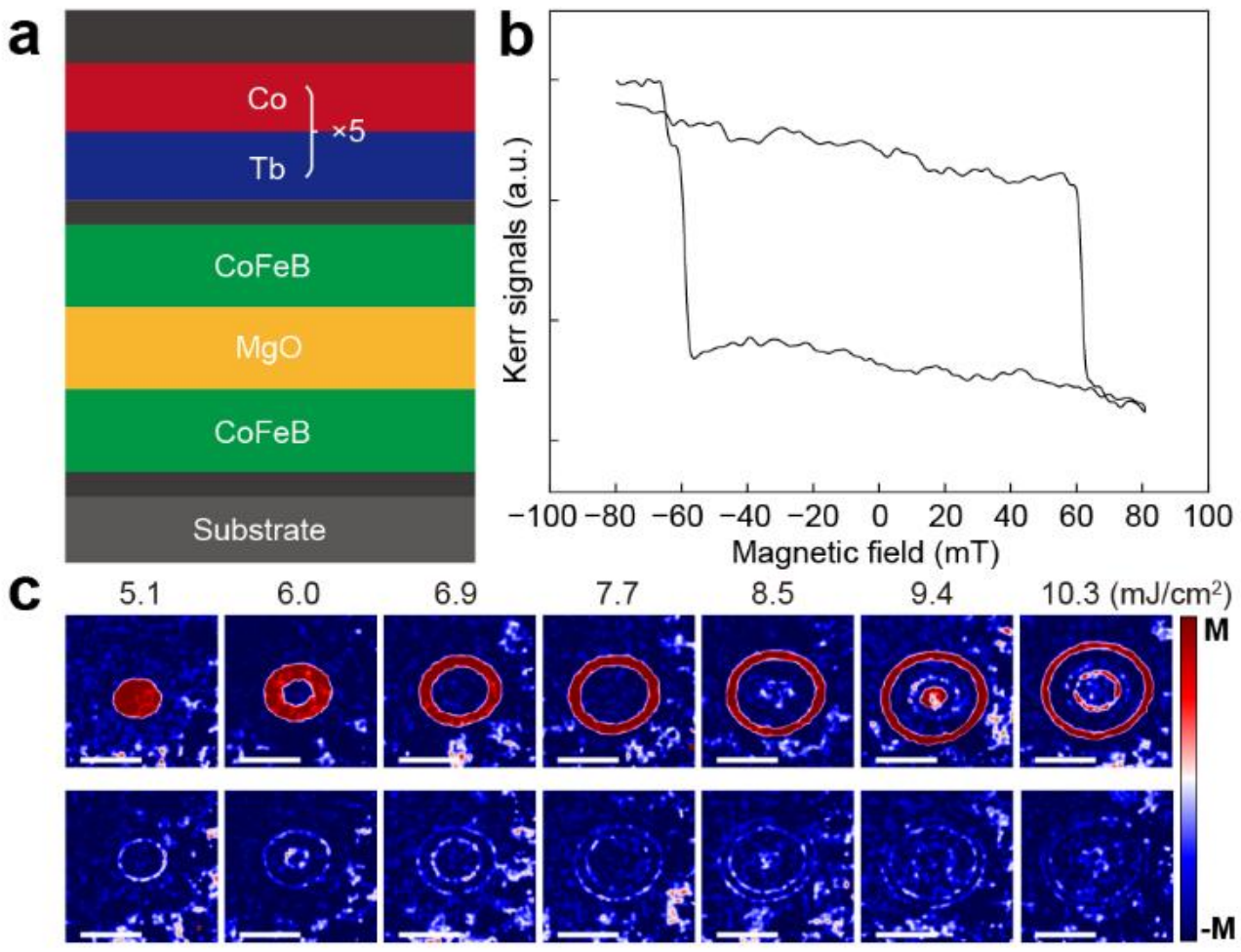

**FIG. S1.** (a) Schematic for the sample structure. (b) The out-of-plane hysteresis curve measured via Kerr microscopy by averaging the signals within regions of interest in MOKE images. (c) The first row shows switched patterns after single-pulse excitation under different fluences. The second row shows the corresponding results by two pulses with a time separation of 2 ms. The scale bars are 50 μm.

## 2. Parameters for macrospin simulation

The numerical simulation performed using a macrospin model consists of solving the Landau-Lifshitz-Gilbert (LLG) equation. The uniaxial magnetocrystalline anisotropy and shape anisotropy contribute to $\mathbf{B}_{\text{eff}}$. The anisotropy easy-axis $\boldsymbol{u}_{\text{k}}$ is tilted by a small angle $\theta_{\text{u}}$ from normal to the sample plane. The Zeeman term is taken into account if external fields are applied. We assume that $K_{\text{u}}(M)$ follows Callen-Callen law, which is proportional to $M^3$. To capture the time evolution for $K_{\text{u}}(M)$, a microscopic three-temperature model (M3TM) is introduced. It describes the energy and angular momentum transfer between electrons, lattices, and spins where the energy input $P$(t) by a single ultrafast laser pulse is given as:

$$P(t) = \frac{\kappa F}{\tau\sqrt{2\pi}l} e^{-(\frac{t-t_0}{2\tau})^2}$$

where $\kappa$, $\tau$, $F$, $l$ correspond to the absorption ratio, pulse duration at standard deviation, fluence and penetration depth. The constant $1/\sqrt{2\pi}$ originates from the energy conservation as an integral over time. For double-pulse simulation, a second pulse is added into $P(t)$. Regarding the spatial profile, the laser spot is assumed to follow a Gaussian distribution, and a factor of ln(2) should be considered to convert the average fluence, evaluated using the FWHM radius, into the corresponding peak fluence at the beam center. The absorption ratio $\kappa$ in the $[\text{Tb}(0.6)/\text{Co}(1.4)]_5$/MTJ film stack is estimated to be ~8.33% by the transfer matrix method [3] as shown in **Fig. S2**. The $l$ is the penetration depth that equals the total thickness of the film 18.4 nm.

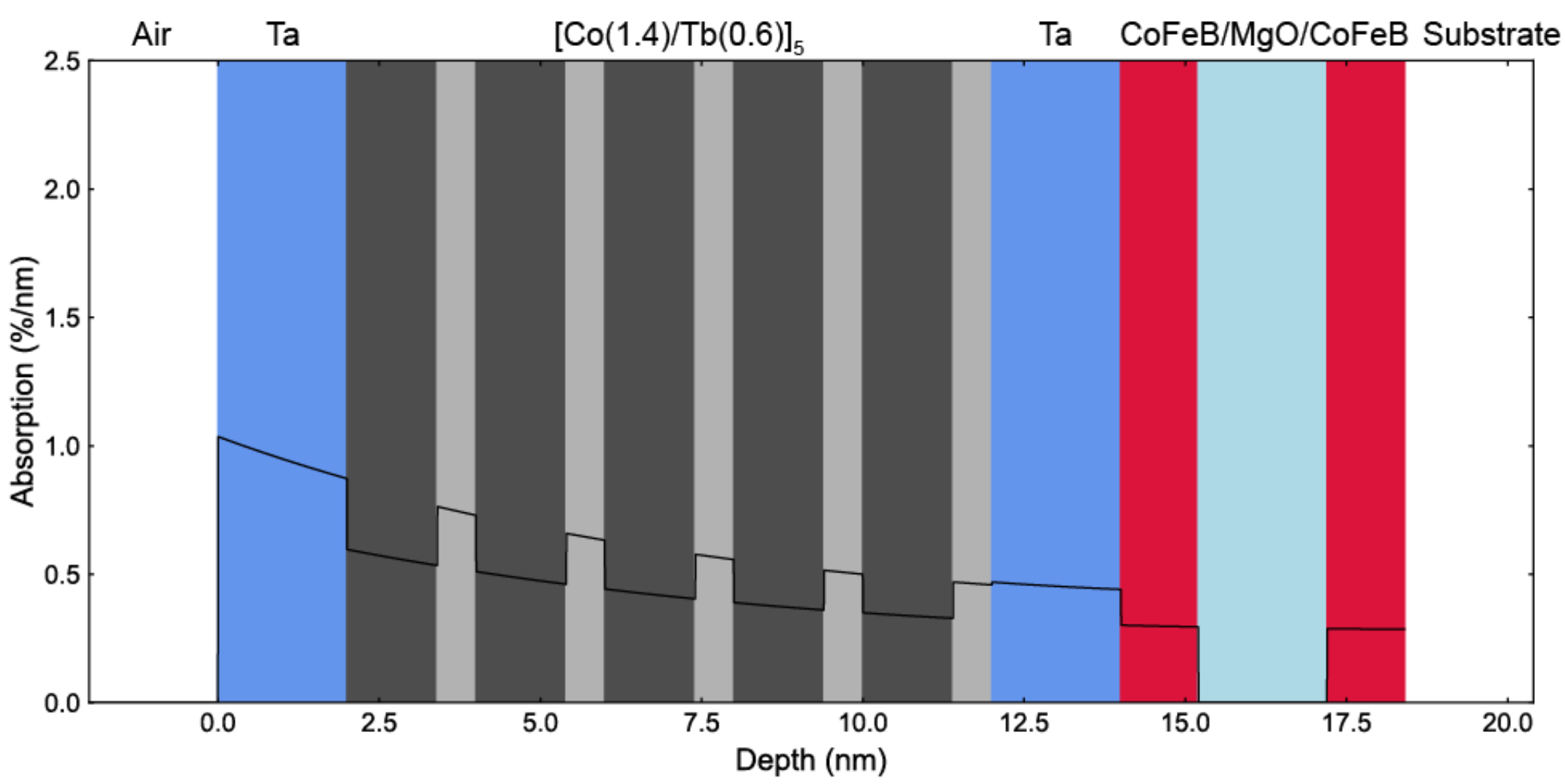

**FIG. S2.** The absorption profile for the $[\text{Tb}(0.6)/\text{Co}(1.4)]_5$ multilayer. The refractive index and extinction ratio for different layers are estimated from refractiveindex.info [4] and [5].

The single-pulse switching behavior is simulated by varying fluences and $\theta_{\text{u}}$ (**Fig. S3**). The images were obtained at equilibrium states after laser excitation. With all $\theta_{\text{u}}$, the switching demonstrates the same trends, except differences in the switching fluence thresholds that decrease at smaller $\theta_u$, consistent with a previous study [6]. As the simulation shows similar trends with experimental results in **Fig. S1**, $\theta_u = 3°$ is chosen as the parameter. All the parameters for M3TM-LLG simulation are listed in **Table S1**.

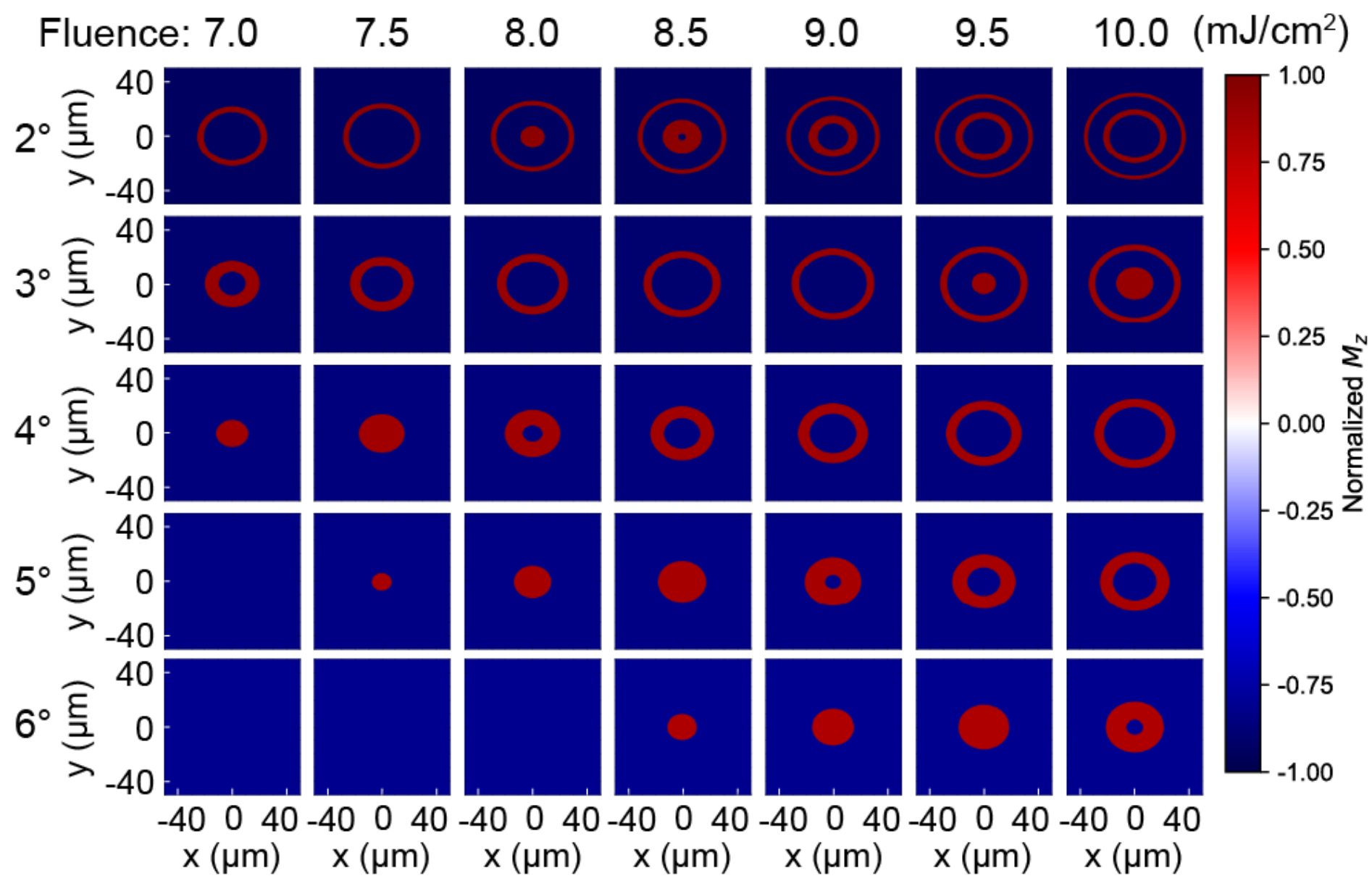

**FIG. S3.** The single pulse switching in the $[Tb(0.6)/Co(1.4)]_5$ multilayer at different fluences and $\theta_u$. The fluences increase from left to right, same as $\theta_u$ from top to bottom. The fluence dependence at $\theta_u = 3°$ is comparable to experimental results.

**Table S1**. Parameters for LLG-TTM model

| Parameter | Value |
|---|---|
| Heat capacity of electrons $C_e = \gamma_e T_e$ | $\gamma_e = 700$ J m$^{-3}$K$^{-2}$ [7] |
| Heat capacity of lattices $C_l$ | $3.0 \times 10^6$ J m$^{-3}$K$^{-1}$ [7] |
| Coupling constant $G$ | $17.0 \times 10^{17}$ J m$^{-3}$K$^{-1}$s$^{-1}$ [7] |
| Pulse duration τ (FWHM) | 500 fs |
| Initial temperature $T_0$ | 300 K |
| Cooling constant $\tau_r$ | 1500 ps [8] |
| Demagnetization rate $R$ | 5 ps$^{-1}$ |
| Saturation magnetization $M_{s,300K}$ | $7.6 \times 10^5$ A m$^{-1}$ [6] |
| Gilbert damping α | 0.1 [6] |
| Gyromagnetic ratio γ | $1.76 \times 10^{11}$ rad T$^{-1}$s$^{-1}$ |
| Curie temperature $T_c$ | 450 K [6] |
| Anisotropy $K_{u,300K}$ | $1.1 \times (\mu_0 M_{s,300K}{}^2/2)$ [6] |
| Tilting angle $\theta_u$ | 3° |

## 3. Double-pulse simulation phase diagram

The simulated double-pulse phase diagram is presented in **Fig. S4**. In the double-pulse simulation, two pump pulses are set to be identical in size (38 × 32 μm$^2$ in FWHM radius) and fluence. The simulation shows good consistency with experimental results in Fig. 2.

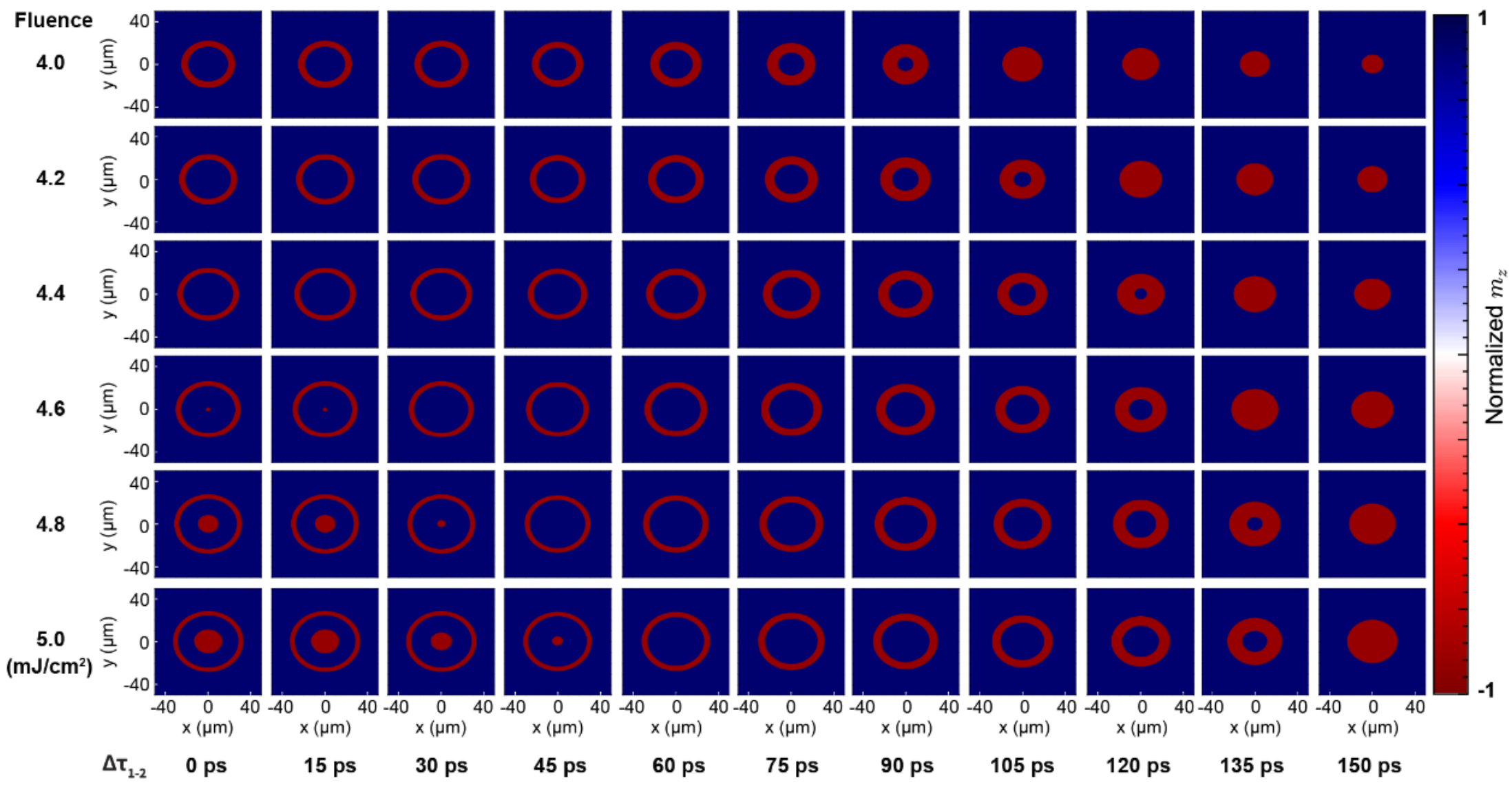

**FIG. S4.** The double-pulse switching phase diagram in the $[Tb(0.6)/Co(1.4)]_5$ multilayer at different fluences and $\Delta\tau_{1-2}$ obtained by macrospin modeling.

## 4. Dynamics of $M$(t) and $K_u$(t)

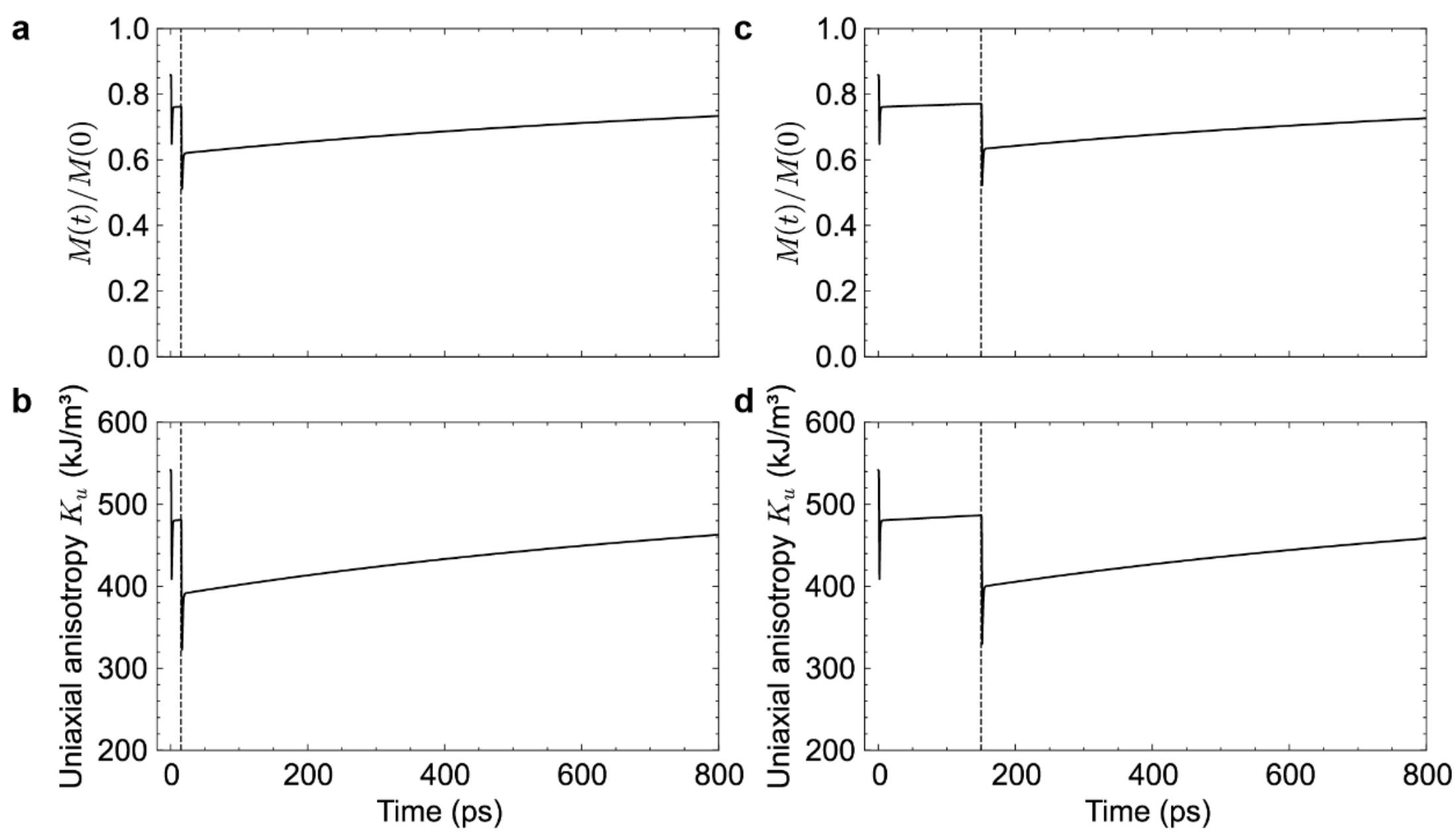

**FIG. S5**. Dynamical evolution under double-pulse excitation with a fluence of 4.4 mJ/cm² and pulse separations of $\Delta\tau_{1-2}$ = 15 ps (a,b) and 150 ps (c,d). Panels (a) and (c) show $M$(t)/$M$(0), while (b) and (d) present the uniaxial magnetocrystalline anisotropy $K_u$. The vertical dash line indicates the timing of the second pulse.

## 5. In-plane field dependence for single-pulse switching

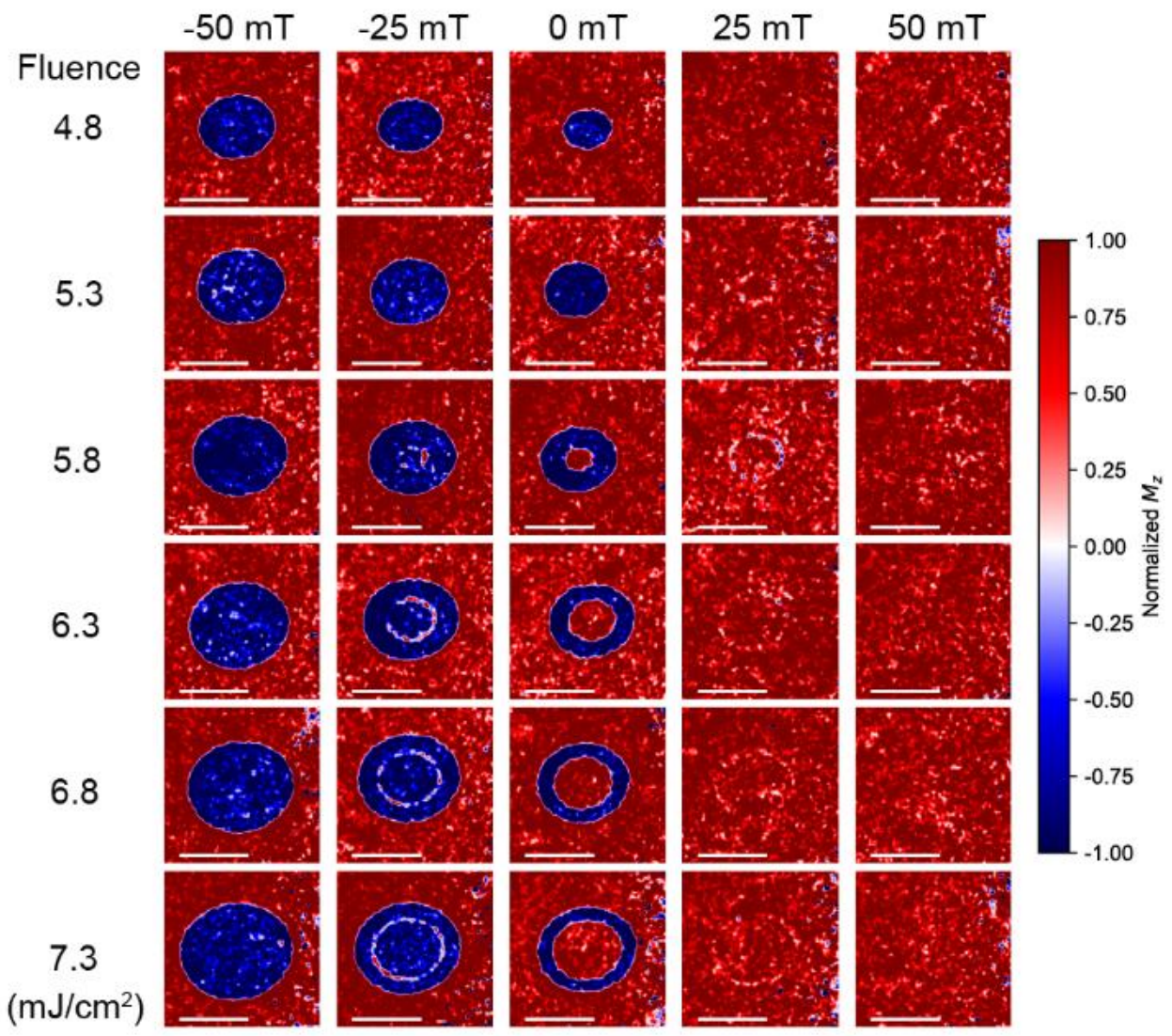

**FIG. S6**. Experimental results of single-pulse switching under in-plane fields.

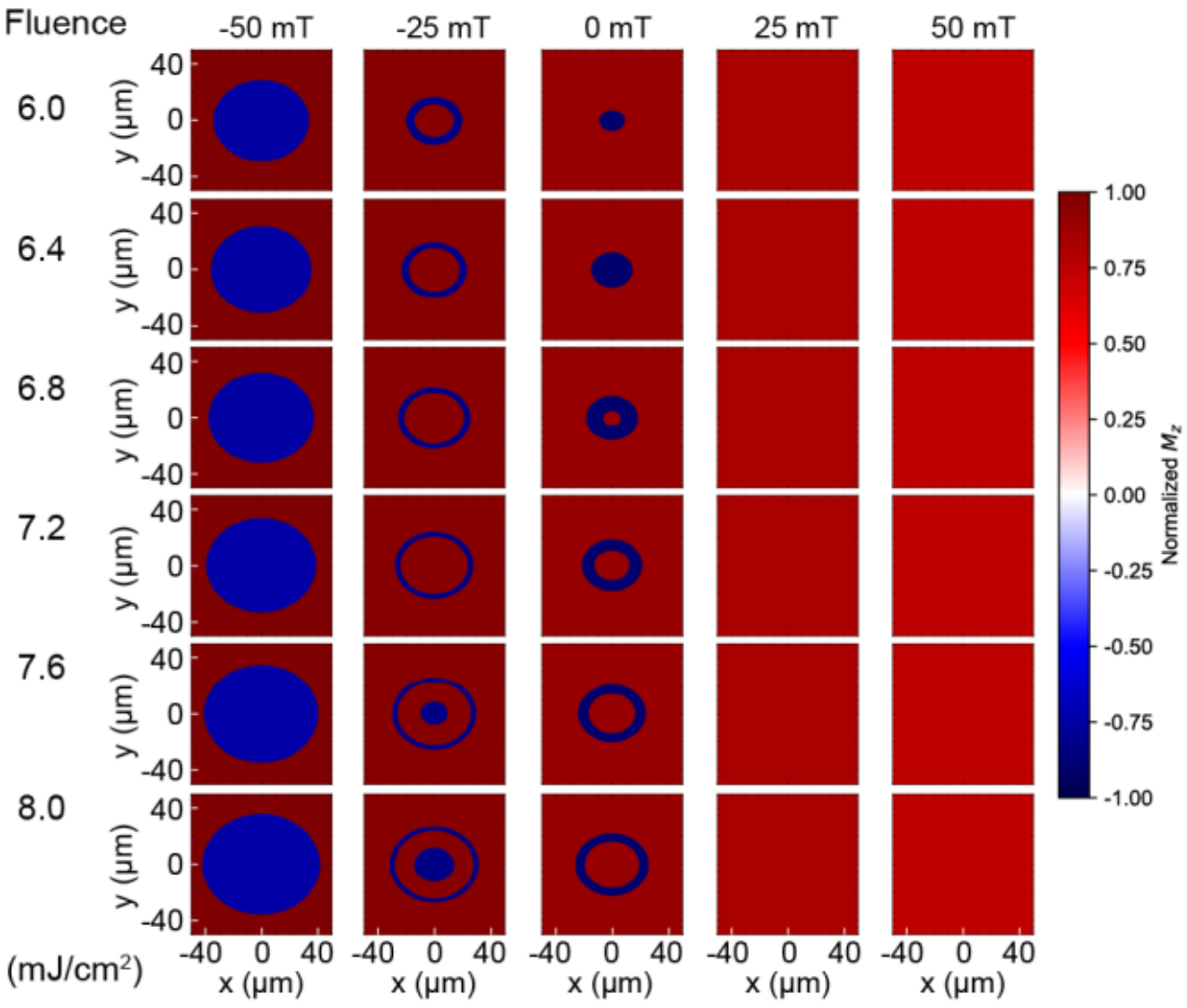

**FIG. S7**. Simulation results of single-pulse switching under in-plane fields.

In **Fig. S6**, the single-pulse excitation is carried out with the existence of an in-plane magnetic bias field in an arbitrary direction. The switching exhibited an enhancement under negative fields, but became suppressed under positive fields. This can be attributed to changes in magnetization canting with bias fields.

The corresponding simulation is demonstrated in **Fig. S7** for comparison. Note that, in contrast to previous modeling approaches, a Zeeman term is included in the effective field $\mathbf{B}_{\mathrm{eff}}$, in addition to contributions from uniaxial magnetocrystalline anisotropy and shape anisotropy. The field is in-plane and aligned with the projection of the tilted axis. The experiments and simulation show generally the same trend. The combination of experiments and simulation provides some evidence for the assumption of a tilted easy-axis.